\documentclass[preprint,english,aps,prb,groupedaddress,floatfix]{revtex4}
\usepackage[T1]{fontenc}
\usepackage[latin9]{inputenc}
\usepackage{amsbsy}
\usepackage{graphicx}
\usepackage{amssymb}
\makeatletter
\usepackage{dcolumn}
\usepackage{bm}
\makeatother
\usepackage{babel}

\long\def\symbolfootnote[#1]#2{\begingroup%
\def\thefootnote{\fnsymbol{footnote}}\footnote[#1]{#2}\endgroup}

\makeatother

\usepackage{babel}

\begin{document}

\title{Reversible magnetism switching in graphene-based systems 
via the decoration of photochromic molecules}

\author{Argo Nurbawono$^{1}$,Chun Zhang$^{1,2}$}
\email{phyzc@nus.edu.sg}

\affiliation{$^{1}$Department of Physics and Graphene Research Centre,National University of Singapore, 2
Science Drive 3, Singapore, 117542\\
  $^{2}$Department of Chemistry, National University of Singapore, 3 Science Drive 3, Singapore 117543}

\date{\today}

\begin{abstract}
By first principles calculations, we demonstrate that when decorated with 
photochromic molecules, it is possible to use light to reversibly control 
the magnetic properties of a nanoscale magnetic system. The combination of a 
graphene-based magnetic system and a photochromic azobenzene molecule is 
chosen as a model system. The {\it trans} and {\it cis} 
isomers of the azobenzene molecule that can be converted between each other 
by means of photoexcitations are found to have drastically different 
effects on the magnetic properties of the system. The results may pave the way for the future design of light 
controllable molecular-scale spintronic devices.
\end{abstract}

\maketitle

Tuning magnetic properties of a material using controllable external methods has 
been a topic of lasting interest in condensed matter physics and also 
material science, which has important applications in the field of 
spintronics. For molecular scale spintronics devices such as tunneling magneto-resistance (TMR) junctions~\cite{CZ, ZX}, spin filters~\cite{Park, ZM} and spin qubits~\cite{Thomas}, devices of interest often have ultra small magnetic centers with the size in the order of one nanometer. It is therefore 
much more challenging to find practical and reliable ways to control the 
magnetic properties of these systems.  
In the current study, via first principles calculations, we propose that with 
the decoration of photochromic molecules, it is possible to use light as an 
effective external control for the magnetic properties of nanoscale magnetic 
systems. The combination of transition metal (TM) embedded graphene and the 
photochromic azobenzene molecule is chosen as the model system for this 
purpose. We expect our computational studies to stimulate new experiments testing the system we proposed and/or searching for other light-controllable magnetic systems.     
     
The discovery of graphene has spurred vast and growing literatures 
for the past several years~\cite{Review}. In this context, there is also 
a persistent interest to investigate and manipulate magnetic properties 
of graphene based systems through various means. We are particularly 
interested in the recent proposal of introducing strongly localized magnetic 
moments into graphene by embedding TM elements in graphene 
vacancies~\cite{Nieminen}. Theoretically, it has been predicted that a range of 
TM elements embedded in graphene can have magnetic moment from 1 $\mu_B$ up to 
more than 3 $\mu_B$ and these magnetic moments are strongly localized around 
these embedded TM elements forming molecular scale magnetic 
centers~\cite{Nieminen,Santos1,Santos2,Kang}. Experimentally Rodriguez-Manzo et al. 
demonstrated that scanning transmission electron microscope (STEM) can be used 
to both image and create vacancies in graphene, which can then be used to 
deposit metal atoms such as Fe, Ni, Co and 
Mo~\cite{ACSNano_vacancy,NanoLett_drillhole}. 
By tailoring the temperatures and concentration of the metal atoms, 
mobility can be controlled to ensure filling of the vacancies.
In most cases the vacancies created by the STEM beam are larger than 
single vacancies (SV) due to the irradiation time and specimen drift 
during irradiation~\cite{ACSNano_vacancy}. This observation was also 
confirmed by Gan et al. who showed that most vacancies in graphene created by 
ion beams are double or multiple~\cite{Small}. Wang et al. have also 
successfully doped a range of metal elements (Pt, Co, Mn) into graphene 
vacancies created with Au and B ion bombardments~\cite{singleatomsubstitution}. 
They noted that there was no contaminations of the ions after the bombardment 
process, and in turn the vacancies can be doped with the desired metal 
elements. Following these successful works, we chose the graphene with Co 
embedded in double vacancies (DV) (Fig. 1) for current 
study since it has been successfully fabricated in different experiments with 
different techniques.

An azobenzene molecule which basically consists of two benzene rings and two 
bridging nitrogen atoms, is a photochromic molecule that can transform 
reversibly between two isomers, namely the {\it trans} 
(Fig. 1(a)) and {\it cis} (Fig. 1(b)) 
isomers upon photo excitation. The energy of the 
{\it trans} isomer is 0.6 eV lower than the {\it cis} isomer~\cite{azo_energy}, 
and the two isomers are separated by an energy barrier of around 1.6 
eV~\cite{azo_barrier}. Both of them are quite stable under room 
temperature. Transformation from {\it cis} to {\it trans} can be induced with 
a laser wavelength of 420 nm, and the opposite way with a laser 
wavelength of 365 nm~\cite{azo_laser_freq}. 
The azobenzene derivatives have been proposed in literature to be used as the basis 
for opto-mechanical devices that can convert the energy of photons to mechanical 
work~\cite{azo_laser_freq} and opto-electronic devices such as light driven 
molecular switches that can switch on and off electrical current in a 
circuit~\cite{Zhang2004,Zhang2006,azo_Nature2007}. A recent work showed that azobenzene molecules have significant effects on spin polarization of organic-ferromagnetic interfaces.~\cite{azofe}  
In this paper, we show by computational studies
that it is possible to effectively tune the magnetic properties of a nanoscale system by the decoration of a light-sensitive azobenzene based molecule.

%\section{R\lowercase{esults and discussions}}

The first principles calculations were performed using density functional 
theory from the Quantum Espresso package~\cite{PWSCF} and we employed 
generalized gradient approximation (GGA) of Perdew-Burke-Ernzerhof (PBE) format 
for the exchange and correlation functionals~\cite{GGA_PBE}. The nuclei and 
frozen core electrons were modeled with norm conserving ultrasoft 
pseudopotentials. Kinetic energy cut off was set above 500 eV and each supercell 
was constructred from $7\times 7$ graphene unit cell with vacuum space more 
than 23 $\text{\text{\AA}}$ between graphene layers. The k-point sampling on the 
graphene plane was $6\times 6$, and this was found to be sufficient and 
excellently reproduce prior results for electronic structures and magnetic 
moments in similar systems with plane-wave basis sets and different choices of GGA functionals~\cite{Nieminen,Kang}.
The structure optimization process was performed with force convergence criteria 
at 0.01 eV/$\text{\AA}$. Spin was relaxed in all parts of the calculations. 

For Co embedded DV graphene, our calculations gave the C-Co bond length around 1.94 $\text{\AA}$ 
and magnetic moment 1.3 $\mu_B$ that are almost exactly the same as literature 
results~\cite{Nieminen}. %The isosurface of the spin density of this system is 
%also plotted in Fig. \ref{SeparateSystems}(a). 
Previous results have shown that the magnetic moment comes from one nonbonding $d$ electron of the Co 
atom~\cite{Nieminen}. Detailed analysis suggested that Co $d_{z^2}$ and $d_{xz}$ 
orbital are almost degenerate near Fermi surface. Only one 
spin channel of the $d_{xz}$ orbital near Fermi surface is occupied, leading to 
the close-to-one $\mu_B$ of magnetic moment. As to azobenzene molecules, our 
calculations reproduces the geometrical structures in good agreement with prior 
calculations both in terms of bond lengths and angles to within 
1\% accuracy~\cite{JACS_azo}.  

\begin{figure}%--------------------------------------------------------- azo dos
\includegraphics[width=10.0cm]{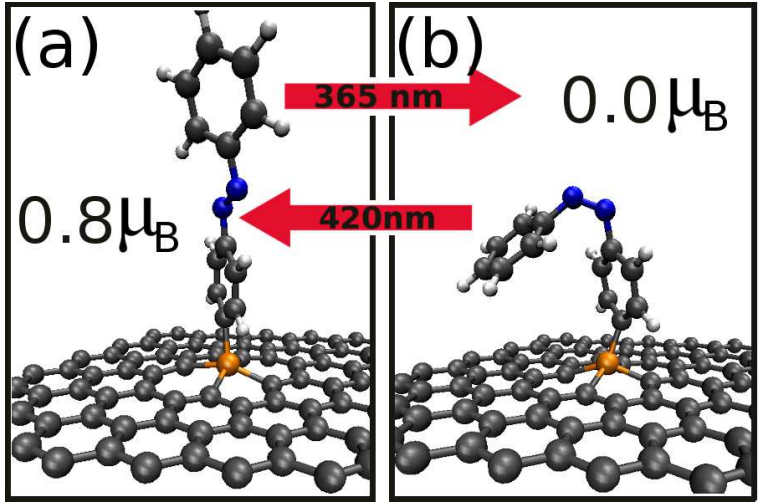}%
\caption{Optimized structures of Co embedded DV graphene 
with (a) {\it trans} and (b) {\it cis} azobenzene molecule adsorbed. We first 
optimized the {\it trans} configuration to obtain the structure as shown in (a). 
The most stable {\it cis} configuration was obtained as follows: First, 
we bent the azobenzene molecule in (a) to be close to its {\it cis} isomer, and 
then relax all degrees of freedom to minimize the total energy. The {\it trans} 
configuration has a magnetic moment of 0.8 $\mu_B$, and the {\it cis} 
configuration is non-magnetic. The two configurations can transform reversibly 
into each other upon exposures of laser beam with two different wavelengths as 
shown in the figure: {\it trans} isomer transforms into {\it cis} by laser 
wavelength $\lambda=365$ nm, and the reverse by $\lambda = 420$ nm.
\label{graphene_layout}}
\end{figure}%-------------------------------------------------------------------

We then consider the adsorption of azobenzene molecule on the Co embedded 
graphene. In order to provide a good contact, one H atom at the end of the 
azobenzene is removed. Since the {\it trans} isomer is the energetic ground 
state, we start from the optimization of the adsorption of {\it trans} 
azobenzene. The adsorptions on the Co and also surrounding carbon atoms were tested.  
It was found that the top site of the Co atom is the mostly preferred. The bonding with the Co atom is more stable by at least 1.2 eV than the bonding with carbon atoms. The most energetically 
favored adsorption configuration is shown in Fig. \ref{graphene_layout}(a). 
The C-Co bond length between the molecule and the Co atom
in this case is about 1.92 $\text{\AA}$, and the adsorption energy of the 
molecule is 2.1 eV. The system with this configuration has a magnetic moment of 
0.8 $\mu_B$, a bit less than the moment without the azobenzene molecule. After 
the adsorption configuration of {\it trans} isomer is obtained, we bend the 
upper benzene ring of the molecule to be close to the {\it cis} isomer, and 
then relax all degrees of freedom to minimize the total energy to determine 
the adsorption configuration of the {\it cis} isomer, which is actually a 
process mimicking the photon induced transformation of the molecule. Note that 
after the optimization, two benzene rings are twisted. The optimized {\it cis} 
adsorption geometry is shown in Fig.  \ref{graphene_layout}(b), where the C-Co 
bond length for {\it cis} is about 1.87 $\text{\AA}$, and the adsorption energy 
2.2 eV similar to the {\it trans} case. %After adsorption, the barrier between two isomers is calculated to be 0.4 $eV$ from {\it cis} to {\it trans} and 1.0 $eV$ from {\it trans} to {\it cis}. 
Very interestingly, with the {\it cis} 
molecule, the system is non-magnetic. 
The drastic change of the magnetic moment from 0.8 $\mu_B$ for {\it trans} 
adsorption to 0.0 $\mu_B$ for {\it cis} case suggests a reversible light 
controllable magnetism switching in the system under study.

\begin{figure}
\includegraphics[width=10.0cm]{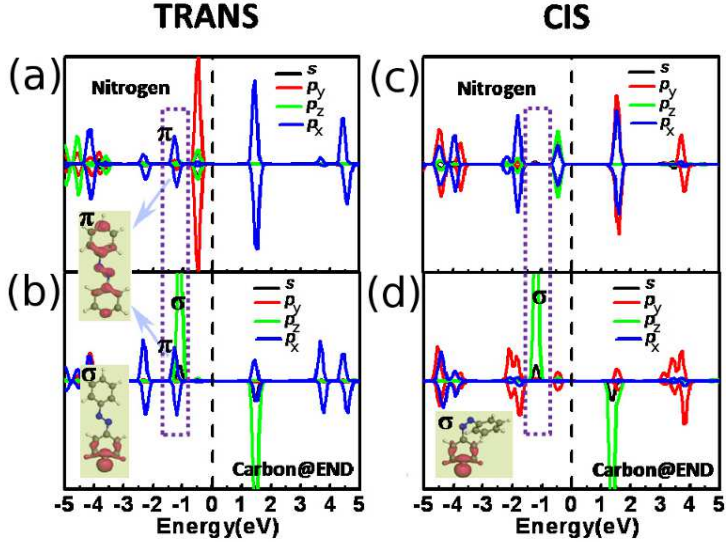}%
\caption{Different electronic structures of two azobenzene isomers after 
removing one H atom at end. Spin-polarized density of states of {\it trans} 
isomer projected on (a) briding nitrogen atoms and (b) end carbon atom without 
hydrogen, and the same for {\it cis} configuration in (c) and (d). The 
isosurfaces of charge density of $\sigma$ and $\pi$-orbital around HOMO are 
shown in insets respectively. Note that for {\it trans} case, there is a $\pi$ 
character orbital that is energetically degenerate with the dangling HOMO-like 
$\sigma$ orbital, while in {\it cis}, this orbital is missing.
\label{PDOS}}
\end{figure}%-------------------------------------------------------------------

The magnetism switching is caused by different electronic 
structures near Fermi surface of two azobenzene isomers, which in turn lead to 
different interactions between two isomers and the magnetic center, the Co atom 
in the surface. To see this more clearly, we calculated electronic structures of 
both azobenzene isomers in gas phase and plotted in Fig. \ref{PDOS} the partial 
density of states (PDOS) projected onto the bridging N atoms and the end carbon 
atom of the molecule that forms bond with the Co after adsorption. As one H atom 
is removed, one dangling bond of the end C atom, $\sigma$ orbital, appears as 
the highest occupied molecular orbital (HOMO) in both {\it cis} and {\it trans} configurations, occupied 
by only one electron. For {\it trans} there is another $\pi$-character orbital 
energetically degenerate with the $\sigma$ orbital (Fig. \ref{PDOS}(b)). This 
'additional' orbital degenerate with HOMO-like $\sigma$ orbital is attributed 
to the linear configuration of {\it trans} isomer in which the $\pi$-orbital of 
N-N bond (Fig. \ref{PDOS}(a)) is delocalized and conjugated with $\pi$-orbital 
of both benzene rings that can be seen from the isosurface of charge density of 
this orbital in the inset of the figure. This conjugated $\pi$-orbital is 
missing in {\it cis} isomer due to the bended form of the molecule as we can 
see from Fig. \ref{PDOS}(c) and (d). This difference between {\it trans} and 
{\it cis} around HOMO at the end-carbon (without H) is crucial for magnetic 
properties of the azobenzene-decorated system, which will be discussed later.

When the azobenzene molecule adsorbed on Co embedded DV graphene, the 
half-occupied $\sigma$-orbital of both {\it trans} and {\it cis} form $\sigma$ 
bond with Co atom. This bond is strong enough to stabilize this hybrid system. 
%Although $\sigma$ bonds in both azobenzene isomers are similar, magnetic 
%properties of {\it trans}, 0.8 $\mu_B$/Co atom, is totally different from that 
%of {\it cis}(non-magnetic). 
The zero magnetic moment of the {\it cis} 
configuration can be understood as the fact that after forming a bond with the 
{\it cis} isomer, the originally nonbonding $d$ orbital of the Co atom is 
satisfied, resulting in a non-magnetic system. The origin of the magnetic moment 
of the {\it trans} configuration is the abovementioned HOMO-degenerate 
$\pi$-orbital that only exists in {\it trans} isomer and the hybridization of 
Co $d$ orbital and C $p_z$ orbital of {\it trans} isomer. 

%\begin{center}
%\includegraphics[width=0.8\textwidth]{Fig4.eps} 
%\par\end{center}

\begin{figure}
\includegraphics[width=10.0cm]{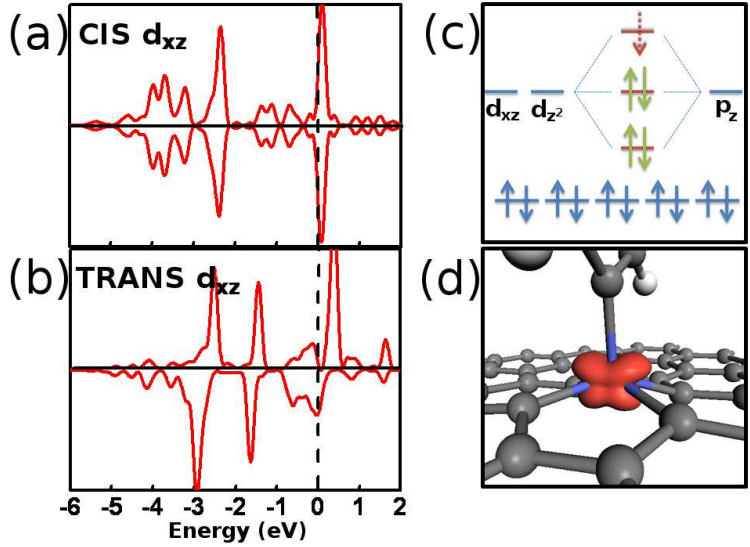}%
\caption{Spin-polarized PDOS of cobalt projected on $d_{xz}$ of (a) {\it cis} 
and (b) {\it trans} configurations. In both cases, the $d_{xz}$ orbital split 
into three peaks after adsorption. Different occupation of the $d_{xz}$ near 
the Fermi surface lead to different magnetic properties of two configurations. 
(c) Schematic demonstration of the hybridization of Co $d$ orbitals near Fermi 
surface with $p_z$ orbital of end carbon in azobenzene molecule. Five occupied 
$\sigma$ orbitals are also shown. After forming these $\sigma$ bonds, the Co 
atom has 4 leftover valence electrons. For {\it cis} case, these four valence 
electrons occupy two low-lying hybridized orbitals, leaving zero nonbonded 
electrons and zero magnetic moment. For {\it trans} case, electrons in the 
HOMO-degenerate $\pi$ orbital of azobenzene populates to the highest hybridized 
orbital (the dashed arrow), resulting in a close-to-one $\mu_B$ of magnetic 
moment. (d) Isosurface of spin density of the system decorated by {\it trans} 
azobenzene. Note the $d_{xz}$ characteristics of the spin 
density. \label{SPLITTING}}
\end{figure}%-------------------------------------------------------------------

When the Co atom is bonded with azobenzene, the $d_{z^2}$ and $d_{xz}$ orbital 
near the Fermi surface hybridize with the $p_z$ orbital of end Carbon atom of 
the molecule, forming three hybridized orbitals as shown in 
Fig. \ref{SPLITTING}(c). The Co atom has nine valence electrons. For {\it cis} 
case, one can assume that five of them go into five $\sigma$ C-Co covalent bonds 
while the left four electrons occupy two lower hybridized states formed by Co 
$d$ and C $p_z$ orbital as shown in Fig. \ref{SPLITTING}(c). All valencies are 
saturated and the magnetic moment is zero. However, for {\it trans} isomer, as 
shown in Fig. \ref{PDOS}, there is an additional $\pi$ orbital degenerate with 
HOMO-like $\sigma$ orbital. Electrons of that orbital will 
also populate to hybridized states (see dashed arrow in Fig. \ref{SPLITTING}(c)) 
that induces a close-to-one $\mu_B$ of magnetic moment. This picture can be 
easily verified by detailed analysis of electronic structures and magnetic 
properties. From the isosurface of spin density of the {\it trans} configuration 
(Fig. \ref{SPLITTING}(d)), we can see that the magnetic moment localized on the 
Co atom mainly has characteristics of $d_{xz}$ orbital, and the PDOS projected 
on Co $d_{xz}$ (Fig. \ref{SPLITTING}(a) and (b)) clearly shows that the orbital 
splits into three orbitals due to the hybridization in both cases. For 
{\it cis}, both spin-up and spin-down orbitals below Fermi energy are equally 
occupied and there is one unoccupied orbital right above the Fermi energy for 
both spin channels, while for {\it trans}, for spin-down channel, there are 
roughly three orbitals occupied, and for spin-up case, only two orbitals 
occupied, which agree with the aforementioned simple picture. %The transfer of the electron from the $\pi$ orbital of the {\it trans} molecule to the hybridized state can be further verified by density of states projected on azobenzene before and after adsorption (Fig. S4 in supplemental information).         

%\section{C\lowercase{onclusions}}

In conclusion, via first principles methods, we analyzed the electronic and 
magnetic properties of a magnetic graphene-based system decorated with the 
photochromic dehydrogenated azobenzene molecule. Two isomers of the azobenzene ({\it trans} 
and {\it cis}) that can be reversibly converted between each other upon photo 
excitation are found to have drastically different effects on magnetic 
properties of the system. %The system with the {\it trans} molecule has a 
%magnetic moment of close-to-one $\mu_B$, and the {\it cis} isomer completely 
%eliminates the magnetic moment. 
These results imply a way to reversibly control 
the magnetic properties of a nanoscale magnetic system using light, which may 
have great implications in future applications of molecular spintronic devices. 
We trust that the similar light-controllable magnetism is can also be found in some other systems, for example, the molecular magnets decorated with photochromic molecules. 

%\begin{acknowledgments}

We acknowledge the support from Ministry of Education (Singapore) and NUS academic research grants (R-144-000-325-112 and R144-000-298-112). Computational works were performed at the Centre for 
Computational Science and Engineering parallel computing facilities.

%\end{acknowledgments}

\vspace*{2\baselineskip}

{\bf{References}}

\end{document}